# Second-Order Non Linear Optical Properties of Zinc Oxide and Aluminum doped Zinc Oxide Thin Films grown by Atomic Layer deposition

Calford O. Otieno
[1]Department of Physics, Kisii University
P.O. Box 408-40200, Kisii, Kenya

In this paper, Second-order NLO properties of ZnO and AZO thin films from experimental results are discussed. Measurements were a single wavelength of 1.064 μm using the standard rotational Maker fringes technique in a transmission scheme. Further broadband dispersion of $\chi^{(2)}$ as characterized by harmonic generation over a broadband wavelength of 1.2-2.1μm are discussed. Experimental results for the *p*-and *s*-polarization of the fundamental beam are presented. In the analysis α-Quartz was used as the standard reference material. All the experiments were conducted at room temperature.

## I. Maker Fringes Theory

The Maker fringes method is a reliable and standard technique to characterize the NLO properties of a material[1,2]. It involves measuring the SHG signal as a function of incident angle for a given fixed polarization state of the input and the SHG beams[3-5]. The measured signal was compared to a standard reference material such as the $LiNO_3$ or crystalline quartz. The Maker fringes technique can be used to study the NLO coefficients of all materials whether the material is phase matching or not. As the sample is rotated, oscillations of SHG power occurs due to phase mismatch. Behavior of the SHG signal and the phase modulation is given by

$$\text{sinc}^2\left(\frac{\Delta k L}{2}\right), \quad (1)$$

where L is the sample thickness and Δk is the wave vector mismatch. For phase matching conditions Δk = 0, but due to the dispersive behavior of the materials, there is always phase mismatch between the input and SHG beam. As the sample is rotated the effective length changes and periodic maxima and minima occur when there is no phase matching[2,6,7]. Figure 1 is an example of a theoretical model of the oscillatory behavior of the SHG Maker fringes pattern of ZnO thin film of thickness L=250nm. The model parameters and equation are from[8,9]. The output SHG count depends on the polarization of the of the input and the output of the beam and the angular position of the sample. The blue trace is for the *p*-polarized beam while the green is for the *s*-polarized beam. As the angle of incidence increases, the SHG counts increases with the local maximum occurring at ~40-45° corresponding to the coherent length, more details, and elaborate theory is reported elsewhere[2]. The oscillatory and angular dependence behavior can be seen to originate from the changing path length resulting into oscillations with increasing angle. Index mismatch and sample thickness are the key factors determining the number of oscillations. The oscillatory behavior may be due to the directionality of the effective second-order nonlinear coefficient $d_{eff}$ is angular dependent and will be discussed in details below.

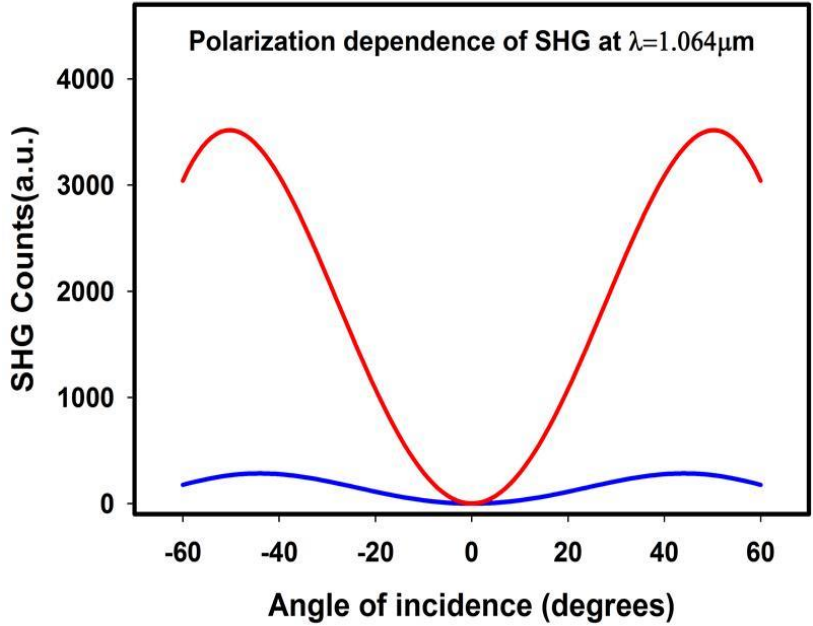

Fig. 1: Theoretical prediction of polarized SHG of 250nm thick ZnO thin film, the blue traces shows the *s*-polarization behavior while the red traces show the *p*-polarization.

## II. Effective nonlinear coefficient $d_{eff}$ for 6mm($C_{6v}$) Crystal Class

In this section, we briefly discuss the effective nonlinear $d_{eff}$ for 6mm($C_{6v}$) crystal class. For maximum and efficient conversion of the fundamental beam to SHG, the phase matching condition and the NLO coefficient are important parameters. The effective NLO coefficient $d_{eff}$ is related to second order susceptibility as $\chi^{(2)}=2d_{eff}$. Consider the coordinate system in Figure 2 below for ordinary and extraordinary rays. The optical axis is along the z-axes. The ordinary polarization is in the xy plane and has $\phi$ dependence and normal to the optics axis and the wavevector. A linear polarized wave can be written as a superposition of extraordinary and ordinary polarization. As shown in the diagram the extraordinary polarization $p^e$ is varying with the azimuthal angle $\phi$ and $\theta$. The nonlinear $d_{eff}$ coefficient can be defined in terms of the unit polarization vectors as [10, 11];

$$\left.\begin{array}{l}p_x^e = \cos\theta\sin\phi\\ p_y^e = \cos\theta\sin\phi\\ p_z^e = -\sin\theta\\ p_{ox} = -\sin\phi\\ p_{oy} = \cos\phi\\ p_{oz} = 0\end{array}\right\}, \qquad (2)$$

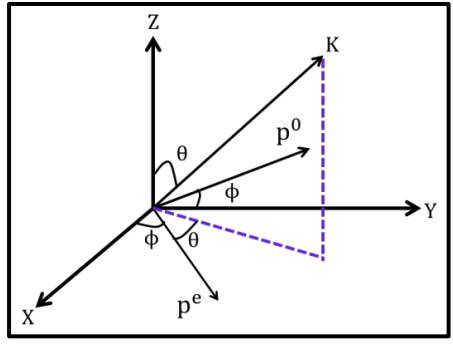

Fig: 2: Coordinate system for extraordinary and ordinary polarization. The ordinary polarization is in the xy plane, and the extraordinary polarization is below the axis. The propagation vector is projected on the xy plane.

Since effective nonlinear coefficient $d_{eff}$ is a third rank tensor, its has twenty seven elements which are reduced by crystal symmetry and Kleinmann conditions[12]. Relevant to this study, we focus on the crystallographic point group 6mm($C_{6v}$) for ZnO and AZO. Nonlinear polarization is related to $d$ coefficient as[13]

$$P_i^{NL} = 2d_{il}E_l^2. \qquad (3)$$

Since $d$ tensor has twenty seven elements we the contracted notation and the intrinsic permutations discussed in chapter, SHG is further simplified 18 elements, the following nonvanishing elements for the 6mm($C_{6v}$) point is readily obtained

$$d = \begin{bmatrix} 0 & 0 & 0 & 0 & d_{15} & 0\\ 0 & 0 & 0 & d_{15} & 0 & 0\\ d_{15} & d_{15} & d_{33} & 0 & 0 & 0 \end{bmatrix}. \qquad (4)$$

The effective nonlinear $d_{eff}$ coefficient is defined in terms of incident beam SHG and unit $p_1$, $p_2$ and $p_3$ as

$$d_{eff} = p_3 \cdot \vec{d} p_1 p_2, \qquad (5)$$

where $p_3$ is the unit vector polarization of the SHG, $p_1$ and $p_2$ are the contribution of the incident field, here $p_{i(1,2,2)}$ has a magnitude of 1. Considering the vector product of the last two terms in the Equation

5, we obtain the matrix representation of the polarization vectors as

$$p_1 p_2 = \begin{pmatrix} p_{1x}p_{2x} \\ p_{1y}p_{2y} \\ p_{1z}p_{2z} \\ p_{1y}p_{2z} + p_{1z}p_{2y} \\ p_{1x}p_{2z} + p_{1z}p_{2x} \\ p_{1x}p_{2y} + p_{1y}p_{2x} \end{pmatrix}, \quad (6)$$

with the elements defined in Equation 2, and the expression of the $d_{eff}$ in Equation 4, the effective nonlinear susceptibility for the s-p polarized beam is written as

$$d_{eff}^{s-p} = (-\sin\phi, \cos\phi, 0) \begin{pmatrix} 0 & 0 & 0 & 0 & d_{15} & 0 \\ 0 & 0 & 0 & d_{15} & 0 & 0 \\ d_{15} & d_{15} & d_{33} & 0 & 0 & 0 \end{pmatrix} \begin{pmatrix} -\sin\phi\cos\phi\cos\theta \\ \cos\phi\sin\phi\cos\theta \\ 0 \\ -\cos\phi\sin\theta \\ \sin\phi\sin\theta \\ (\cos^2\phi - \sin^2\phi)\cos\theta \end{pmatrix}$$

$$= (-\sin\phi, \cos\phi, 0) \begin{pmatrix} -d_{15}\sin\phi\sin\theta \\ d_{15}\cos\phi\sin\theta \\ -d_{15}(-\sin\phi\cos\phi\cos\theta + \cos\phi\sin\phi\cos\theta) \end{pmatrix}$$

$$d_{eff}^{s-p} = -d_{15}\sin\theta. \quad (7)$$

From the results of Equation 7, the nonlinear $d_{eff}$ tensor component $d_{15}$ is explicitly measured. However, for p-in p-out configuration the tensor $d_{eff}^{pp}$ is dependent on dependent on more than one component as illustrated below

$$d_{eff}^{p-p} = (-\cos\theta, 0, \sin\phi) \begin{pmatrix} 0 & 0 & 0 & 0 & d_{15} & 0 \\ 0 & 0 & 0 & d_{15} & 0 & 0 \\ d_{15} & d_{15} & d_{33} & 0 & 0 & 0 \end{pmatrix} \begin{pmatrix} \cos^2\theta \\ 0 \\ \sin^2\theta \\ 0 \\ -2\cos\theta\sin\theta \\ 0 \end{pmatrix}$$

$$d_{eff}^{pp} = (-\cos\theta, 0, \sin\phi) \begin{pmatrix} -d_{15} \; 2\cos\theta\sin\theta \\ 0 \\ d_{15}\cos^2\theta + d_{33}\sin^2\theta \end{pmatrix}$$

$$d_{eff}^{p-p} = d_{15}\cos\theta\sin 2\theta + \sin\phi(d_{15}\cos^2\theta + d_{33}\sin^2\theta). \quad (8)$$

For p-p polarization the $d_{33}$ component is also dependent on $d_{15}$ according to Equation 8. It is necessary to measure both the s- and p-polarization to determine the nonvanishing components $d_{33}$ and $d_{15}$ for wurtzite structures. In our experimental results discussed later in the chapter we show that $d_{33}$ is always the dominant component. extract act $d_{15}$ and $d_{33}$.

Equation (1), (7) and (8) will be used to fit the experimental data and

## III. Experimental Details for SHG Measurement

SHG efficiency is strongly dependent on the crystalline structure. In particular, SHG occurs in materials which are noncentrosymmetric. It is important to note that SHG entirely vanishes when the input beam is aligned along the c-axis of the sample due to the in-plane symmetry. Due to In our measurements, we used a polarizer before the samples and an analyzer after the sample. The polarizer was to select the preferred polarization of the input beam, and the analyzer was to adjust the polarization of the output beam. The samples were placed on a well calibrated rotating stage to allow for angular variations. Our excitation source was a mode-locked Nd; YAG laser (PL2250 series) which supplied a fundamental beam ($\lambda$= 1064nm, 30ps) at 50Hz. The SHG signal from the sample was measured in transmission geometry. The transmitted signal was collected by a fiber-optic bundle and channeled to a charge coupled device (CCD). Extra precaution was taken to ensure was no any surface-induced effect and SHG symmetry SHG measurements are done with a finite angle of incidence. Thus, the best way was to probe SHG response was via angular Maker fringe experiments[1, 2].

Figure 3 shows a schematic diagram of the SHG measurements signals from other optical components. The spectrally resolved SHG signals at $\lambda/2$ was precisely calibrated with the known and measured efficiencies of the system which included the selective gratings, detectors, and fibre bundle. The SHG was measured for s- and p- polarization beam. The fundamental beam was mildly focused on the sample at normal incidence relative to the surface normal to the sample. The sample was rotated along the between $\pm55^{o}$. α-quartz thickness ~645µm was used as a reference to calculate $\chi^{(2)}$.

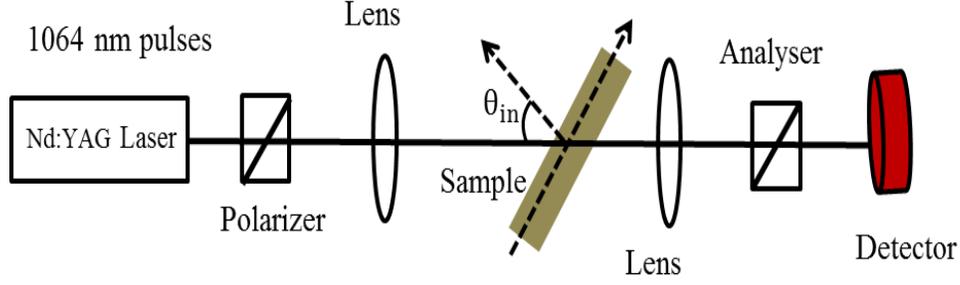

Fig. 3: Experimental setup for SHG Maker fringes measurements[14].

## IV. SHG Results and Analysis

The transmitted intensity $I_{2\omega}$ of SHG considering the angular variations and the air-film-substrate-air model is expressed as[8]

$$I_{2\omega(\theta)} = \frac{128\pi^3}{cA} \frac{(t_{af}^{1\gamma})^4 (t_{fs}^{2p})^2 (t_{sa}^{2p})^2}{(n_{2\omega}\cos\theta_{2\omega})^2} \times I_\omega^2 \left(\frac{2\pi L}{\lambda}\right)^2 \left(\chi_{eff}^{(2)}\right)^2 \frac{\sin^2\phi}{\phi^2}, \qquad (9)$$

where $t_{af}^{1\gamma}, t_{fs}^{2p}$ and $t_{sa}^{2p}$ moreover, are field transmission coefficient of SHG beam from air to film, film to substrate and substrate to air respectively. L is the film thickness and $\chi_{eff}^{(2)}$ is the susceptibility tensor. A is the area of the incident beam spot. $I_\omega$ is the intensity of the incident fundamental beam, $n_\omega$ ($n_{2\omega}$) is the refractive index at the fundamental and SHG frequency inside the film. Represent the transmission coefficient of the fundamental beam from air to the film and ϕ is the coherent length given as

$$\phi = \frac{2\pi L}{\lambda}(n_\omega \cos n_{2\omega}\cos\theta_{2\omega} - n_{2\omega}\cos\theta_{2\omega}), \qquad (10)$$

where $\theta_\omega$ and $\theta_{2\omega}$ are refraction angles in the films determined by

$$\sin\theta = n_\omega \sin\theta_\omega (\sin\theta = n_{2\omega}\sin\theta_{2\omega}). \qquad (11)$$

The effective susceptibility $\chi_{eff}^{(2)}$ in Equation 9 is expressed depending on the polarization state of the fundamental. Wurtzite ZnO and AZO belongs to noncentrosymmetric group SHG[15, 16]. In the absence of the external field, a group theoretical analysis predicts the following independent nonzero elements[17-19]; $\chi_{yzy}^{(2)} = \chi_{xzx}^{(2)}, \chi_{xxz}^{(2)} = \chi_{yyz}^{(2)}$, $\chi_{zxx}^{(2)} = \chi_{zyy}^{(2)}$ and $\chi_{zzz}^{(2)}$. The nonvanishing components depend on the incoming frequency via a dispersion relation[17], the 6mm($C_{6v}$) and, consequently, they give rise to values of nonlinear coefficients are measured in pm/V. For an isotropic and uniaxial media where the average optical axis is along the c-axis, the SHG is always

p- polarized regardless whether the incoming beam is *s*- polarized or *p*-polarized[18, 20, 21]. For s-polarization, we rewrite Equation (7) for the effective nonlinear susceptibility [8, 21]

$$\chi^{(2)}_{eff} = \chi^{(2)}_{xxz}\sin 2\theta_{2\omega}. \quad (12)$$

For p-polarized beam, Equation (8) for the $\chi^{(2)}_{eff}$ is given as[20, 21]

$$\chi^{(2)}_{eff} = \chi^{(2)}_{zxx}(\cos 2\theta_{2\omega}\sin 2\theta_{2\omega} + \sin 2\theta_{2\omega}\cos^2\theta_\omega) + \chi^{(2)}_{zzz}\sin\theta_{2\omega}\sin^2\theta_\omega. \quad (13)$$

Experimental Maker fringe data obtained from ZnO and AZO are presented in Figure 3(a-d) for both s- and p-polarization. The solid lines are the theoretical fits using Equation (1) and (12) for s-polarization and Equation (1) and (13) for p-polarization.

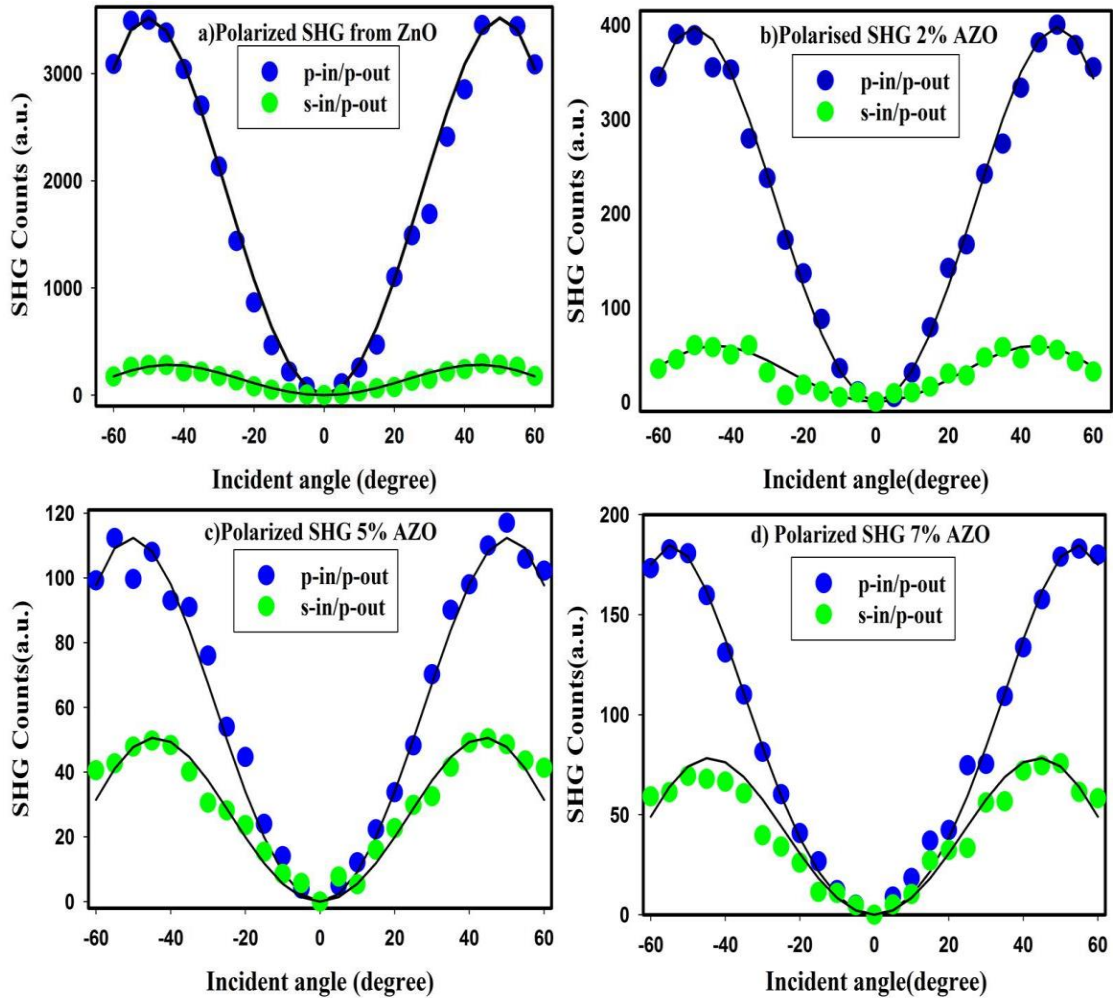

Fig. 4: Polarization dependence of SHG of 250nm thick ZnO and AZO samples measured by Maker fringes in transmission scheme. *s*-polarized (green dots) and *p*-polarization (blue dots), the solid lines correspond to the theoretical fits.

Reasonable symmetry in the SHG counts as a function of incidence angle around the normal incidence implies a high quality of the sample as well as a proper optical alignment. The measured SHG counts increases when the incidence angle was varied from normal incidence and reached a maximum conversion at about $\theta \sim \pm 45^0$, which corresponds to the primary maximum in the oscillatory phase-matching factor. To determine the two nonvanishing $\chi^{(2)}$ components of ZnO, the quantity we first obtained overall constants in the Equation (1) including the (I/A) by fitting the Maker fringe results for quartz. The overall value obtained from the quartz fit is used to fit the ZnO and AZO data. First $\chi^{(2)}_{zxx}$ was independently obtained by fitting Equation (12) and $\chi^{(2)}_{zzz}$ is obtained by fitting the experimental data with Equation (13) above[13]. From the fitting, we found the highest value of $\chi^{(2)}_{zzz}$ ranges from 2.24-12.8pm/V depending on the level of the Al-doping. Our value compares well with the literature values[8, 14, 20]. The lower the doping, the higher the $\chi^{(2)}_{zzz}$ and $\chi^{(2)}_{zxx}$. These results in Table 1 indicate a significant drop in the susceptibility value with Al-doping. This observation is attributed to the decreased crystallinity with increased Al-doping as detailed in the next section. The ratio of the $\chi^{(2)}_{zzz}/\chi^{(2)}_{zxx}$ ranged from 1.14-5.12 with undoped ZnO giving the highest ratio and nearly constant for the doped samples. This ratio provides information about the crystallinity of the sample, for examples for wurtzite-like structures the ratio should be ~2 for good quality samples[22]. Chan et al. found the ration of $\chi^{(2)}_{zzz}/\chi^{(2)}_{zxx}$ to be less than one and attribute the results to poor crystallinity of the nanorods[14]. The absolute values of $\chi^{(2)}_{zzz}$ for ZnO is vastly studied, the outcome depends on among other things, growth kinetics, and the excitation source. In Table 1 we compare values obtained in the literature for the undoped ZnO indicated the absolute value of $\chi^{(2)}_{zzz}$ is varied from different samples with different characteristic (see Table 1 below).

Table 1 NLO coefficients of ZnO and AZO compared with other studies

| Sample | $\chi_{zxx}$(pm/V) | $\chi_{zzz}$(pm/V) | $\left|\frac{\chi_{zzz}}{\chi_{zxx}}\right|$ | References |
|---|---|---|---|---|
| Zinc Oxide | 2.50 | 12.8 | 5.12 | This work |
| 2% AZO | 1.68 | 4.10 | 2.44 | This work |
| 5% AZO | 1.71 | 2.28 | 1.33 | This work |
| 6% AZO | 1.73 | 2.45 | 1.14 | This work |
| 7% AZO | 1.76 | 2.42 | 1.37 | This work |
| 8% AZO | 1.86 | 3.31 | 1.78 | This work |
| Sputtered film | 9.2 | 2.24 | 2.8 | Reference[22] |
| Thin film | - | 12.9 | - | Reference[23] |
| Thin films | 4.6 | 7.0 | 1.52 | Reference[24] |
| Bilayers | 2.6 | 4.2 | 1.61 | Reference |
| Films | 3.6 | 13.4 | 3.77 | Reference[8] |
| Thin film | - | 17.89 | - | Reference[17] |
| nanorods | 2.88 | 18.0 | 0.16 | Reference[14] |
| Nano layers | - | 14.0 | - | Reference[9] |

# V. Doping Dependence of $\chi^{(2)}_{zzz}$ and $\chi^{(2)}_{zxx}$

In this section, we discuss the dependence of $\chi^{(2)}_{zzz}$ and $\chi^{(2)}_{zxx}$ with Al-doping. Figure 5 below shows a decreasing trend of $\chi^{(2)}$ with increased Al-doping. The observed trend is be attributed to; 1) decrease in crystallinity with increased doping. 2) Loss of noncentrosymmetric structure with increased doping. It is known that the efficiency of the SHG depends on the crystallinity and symmetry of the material[25].

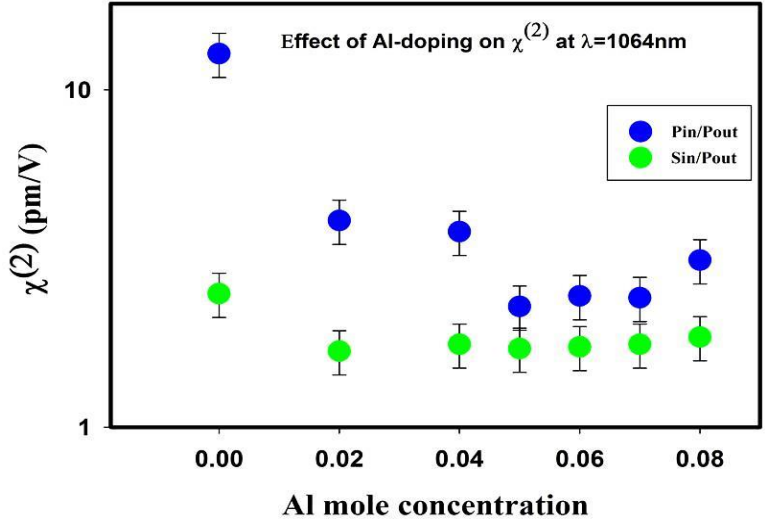

Fig. 5: Dependence of $\chi^{(2)}_{zzz}$ (blue dots) and $\chi^{(2)}_{zxx}$ (green dots) with increased Al-doping.

The possible effects of doping its impact on sample crystallinity and symmetry properties and nonlinearity of the samples include and not limited to;

1) Distortion of the crystal lattice; when $Al^{3+}$ ions with radius~0.053nm substitute $Zn^{2+}$ ions with radius~0.074nm the crystal structure is distorted due to the small size of the $Al^{3+}$ions[24].
2) $Al^{3+}$ dopants have a tendency to segregate at the grain boundaries

resulting to deterioration crystal lattice. Thus significantly reduces SHG conversion because Segregated $Al^{3+}$ at the grain boundaries will inhibit the grain boundary contribution to the SHG due to reflective scattering at the grain boundaries[26, 27].

3) Material self-compensation effect; is the tendency of the crystal to lower its energy by forming defects to counter the dopant atoms. Increased Al-doping may result in the formation of $Zn^{2+}$ compensating defects to counter $Al^{3+}$ dopants. These $Zn^{2+}$ defects may result in decreased SHG efficiency[28].

4) The preferred orientation of ZnO and AZO is along (001) with the optical axis normal to the sample surface, with increased Al-doping the sample suffers loss of the noncentrosymmetric due to the slight shift in the optical axis as the samples tend to be amorphous[24]

5) Wang et al[17] has advanced the concept of stacking defaults. For example, the hexagonal wurtzite structures with c axis orientation may have different stacking sequence during the deposition[17]. Grain may flip to a different stacking sequence resulting into growth in a different axis other than that of the c-axis.

From the theoretical fit, we obtained we ration $\chi^{(2)}_{zzz}/\chi^{(2)}_{zxx}$ for all samples. As reported before this ratio should be ~2 for wurtzite samples of good crystallinity[19, 22, 29]. The ratio of these components provides useful information concerning the degree of crystallinity. In our measurement, the ratio of these components ranged from 1.14-5.12±0.03 and is shown in Table 1. These results indicated the ALD grown ZnO and AZO samples were of good crystallinity. Further crystallinity evidenced by the large $\chi^{(2)}_{zzz}$ measured for ZnO. Our values also compared to the reported literature values.

## VI. Broadband dispersion of $|\chi^{(2)}_{zzz}|$

We conducted a wavelength dependent SHG measurement between $\lambda$=1.2-2.1μm for p-polarized input beam. We limited our measurements of broadband to p-polarization for reasons stated elsewhere[19]. To obtain the broadband $\chi^{(2)}_{zzz}$, used our Maker fringe experiments results at $\lambda$=1.064 μm reported above. Considering the Maker fringe Equation 1 for two other wavelengths, a general expression for the $\chi^{(2)}_{eff}(\lambda)$ at any wavelength is obtained. The general expression was obtained as follows, let the SHG intensity at $\lambda$=1.064 μm be

$$I'_{2\omega(\theta)} = \frac{128\pi^3}{cA} \frac{(t^{1\gamma'}_{af})^4 (t^{2p'}_{fs})^2 (t^{2p'}_{sa})^2}{(n_{2\omega}\cos\theta_{2\omega})^{2'}} \times I'^2_\omega \left(\frac{2\pi L}{\lambda'}\right)^2 \left(\chi^{(2)'}_{eff}\right)^2 \frac{\sin^2\phi'}{\phi^{2'}}, \quad (14)$$

where all symbols bear their usual meaning as discussed above and letting Equation 1 represent any other wavelength, dividing the two Equations results into results into a

general expression for $\chi_{eff}^{(2)}(\lambda)$ at any other wavelength

$$\chi_{eff}^{(2)}(\lambda) = \left(\frac{(A'.I_\omega^2)^{1/2}\lambda}{(A.I_\omega^{2'})^{1/2}\lambda'}\right)\left(\frac{(n_{2\omega}\cos\theta_{2\omega})^2}{(n'_{2\omega}\cos\theta'_{2\omega})^2}\right)\left(\frac{\sin^2\phi'\phi^2}{\sin^2\phi\phi^{2'}}\right)\left(\chi_{eff}^{(2)'}\right), \quad (15)$$

where A represent the field transmission coefficients as defined in Equation (1), $\chi_{eff}^{(2)'}$ is the effective susceptibility of the reference determined at the reference wavelength, $\chi_{eff}^{(2)}(\lambda)$ is the susceptibility at any other wavelength of choice. The primed values are all at the reference wavelength of 1.064μm.

Figure 4 below show the dispersion of $\chi_{zzz}^{(2)}$ as a function of the wavelength. We can infer that Al-doping did not cause any enhancement due to reasons stated in section xx above. This result is consistent with the measurement at 1.064μm. Our experimental range for broadband dispersion i.e. λ/2= 0.6-1.05μm is away from the bandgap of ZnO ~0.368μm so no bandgap enhancement effect was expected. Kulyk et al. reported decreased $\chi^{(2)}$ in ZnO embedded in PMMA[30] and silver doped ZnO[24] attributing their observation to the distorted crystal structure.

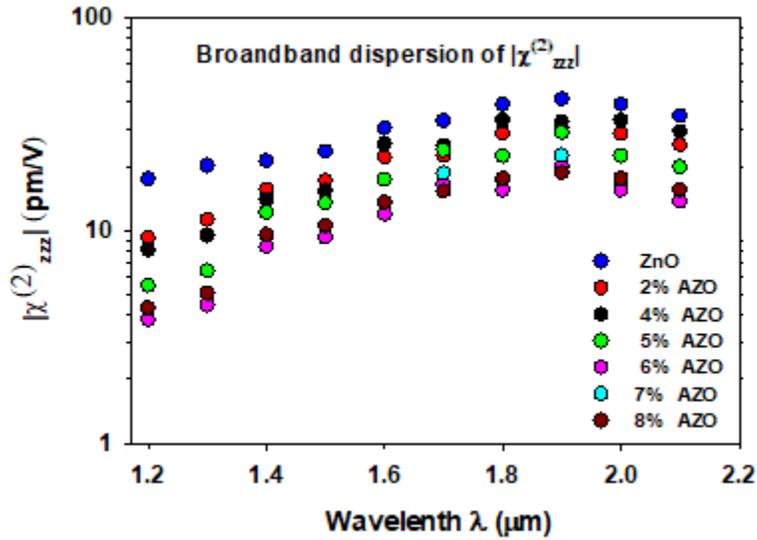

Fig. 6: Wavelength-dependent second-order NLO susceptibility in ZnO and AZO. $\chi^{(2)}$ did not so any enhancement on the Al-doping.

## VII. Comparison of $\chi_{zzz}^{(2)}$ of AZO Thin Film and other X-doped ZnO

Several studies on second-order nonlinearity of doped ZnO have been reported. We provide a comparison of the ZnO, AZO, and other X: ZnO (X: F, Ga In, Al, Co, Ag, and Au). Different authors have reported different values, the slight disparity of the values are due to the different preparation technique and excitation sources.

Table 2. Effects of X-dopant on NLO properties of ZnO

| Deposition technique | Thickness L(nm) | Excitation Source/ λ(nm) | Dopant | $\chi^{(2)}_{zzz}/\chi^{(2)}_{zxx}$ (pm/V) | Reference |
|---|---|---|---|---|---|
| Chemical spray | 550 | τ=5nm,12KHZ λ=1064nm | Flourine | 16.4/5.3 | A[31] |
| Chemical spray | 500 | τ=5nm,25KHZ λ=1064nm | Flourine | 26.4/10.3 | B[32] |
| RF sputtering | 600 − 1200 | τ=16ps,10HZ λ=1064nm | Cu/Ag | 17,13/1.4,1.0 | C[24] |
| Sputtering | 1000 | τ=16ps,10HZ λ=1064nm | PMMA | 1.0-5.1 | D |
| Spray Ultrasonic | 300 | τ=16ps,10HZ λ=1064nm | Nickel | 0.36-0.65 | E |
| **ALD** | **250** | **τ=30ps,50HZ λ=1064nm** | **Aluminum** | **12.8 /2.5** | **This work** |

## VIII. Conclusion

Complete study on the second-order NLO properties of ZnO and AZO prepared by ALD are presented. The magnitude of $\chi^{(2)}_{zzz}$ and $\chi^{(2)}_{zxx}$ measured at 1.064μm are discussed. Our results indicate decrease in magnitude of $\chi^{(2)}_{zzz}$ and $\chi^{(2)}_{zxx}$ with increased Al-doping and we attribute the observation to decrease in crystallinity with increased Al-doping. The dispersion of $\chi^{(2)}_{zzz}$ over a broadband range 1.2-2.1μm indicated result consistent to those at 1.064μm. The SHG measurement confirmed the preferential orientation of grains of the polycrystalline samples were along the c-axis. Our results were in good agreement with the literature values especially the large $\chi^{(2)}_{zzz}$ reported. We propose future studies involve measurement of the SHG with different dopants grown under same conditions for comparison, Density functional theory calculations are proposed.